\documentclass[apj]{emulateapj}

\usepackage{apjfonts}

\shorttitle{Photometric Variability of a T$2.5$ Brown Dwarf}
\shortauthors{Artigau et al.}

\begin{document}

\title{Photometric Variability of the T$2.5$ Brown Dwarf SIMP J013656.5+093347; \\
   Evidence for Evolving Weather Patterns.}

\shorttitle{Photometric Variability of SIMP0136}

\author{\'Etienne Artigau \altaffilmark{1}}
\author{Sandie Bouchard \altaffilmark{2}}
\author{Ren\'e Doyon \altaffilmark{2}}
\author{David Lafreni\`ere \altaffilmark{3}}
\email{eartigau@gemini.edu, bouchard@astro.umontreal.ca, doyon@astro.umontreal.ca, lafreniere@astro.utoronto.ca}

\altaffiltext{1}{Gemini Observatory, Southern Operations Center, Association of Universities for Research in Astronomy, Inc., Casilla 603, La Serena, Chile}
\altaffiltext{2}{D\'epartement de Physique and Observatoire du Mont M\'egantic, Universit\'e de Montr\'eal, C.P. 6128, Succ. Centre-Ville, Montr\'eal, QC, H3C 3J7, Canada}
\altaffiltext{3}{Department of Astronomy and Astrophysics, University of Toronto, 50 St. George Street, Toronto, ON M5S 3H4, Canada}

\begin{abstract}

We report the discovery of a photometric variability in the bright T$2.5$ brown dwarf SIMP J013656.57+093347.3. Continuous $J$-band photometry has been obtained for several hours on four different nights. The light curves show a periodic modulation with a period of $\sim$2.4 hours, a peak-to-peak amplitude of $\sim$ 50 mmag and significant night-to-night evolution. We suggest that the light curve modulation is due to the brown dwarf's rotation and that the longer term variations come from surface features evolution and/or differential rotation. We obtained complementary observations over a single night in the $J$ and $K_{\rm s}$ bands; the object displays correlated photometric variability in both bands, albeit with smaller $K_{\rm s}$-band amplitude. The ratio of the $K_{\rm s}$ and $J$ variability amplitudes puts strong constraints on the physical mechanisms at play. Based on theoretical models of brown dwarf atmospheres, our results suggest that the atmosphere of SIMP0136 is comprised of both grain-free and colder (by $\sim 100$ K) grain-bearing cloudy regions. This discovery, and its interpretation, provide a natural explanation of the so-called $J$-band brightening.

\end{abstract}


\section{Introduction}

The study of brown dwarfs (BD) is a very active field in astronomy. Since the discovery of the first ones less than 15 years ago \citep{Rebolo1995, Nakajima1995}, much has been learned on the nature of those objects. While spectacular advances in modeling their atmospheres have been made, understanding the physical properties of BDs straddling the L/T transition remains one of the most challenging problems. Among the unresolved issues is the intriguing increase of the $J$-band luminosity with decreasing temperature over the 1400-1100 K interval, just following the transition from L to T type. This odd behavior, the so-called $J$-band brightening \citep{Vrba2004}, is surprising as one would naively expect near-infrared brightness to decrease with temperature. The brightening coincides with the disappearance of silicate grains from the brown dwarfs' atmospheres and is most likely  a consequence of this disappearance, but a self consistent explanation has yet to be found. While dusty models containing no grain sedimentation best reproduce the behavior of late-M to mid-L dwarfs, condensed (clear) models, where the grains are entirely removed from the atmospheres, accurately describe mid-to-late-T dwarfs. Models including grain settling and precipitation \citep{Allard2003, Burrows2006} have been able to reproduce a flattening of the $J$-band magnitude over the problematic temperature interval, but no single set of physical parameters can entirely reproduce the observed early-T dwarf sequence. The fundamental assumption of these models, that a single set of physical parameters can describe the photosphere of T dwarfs, might not be accurate. The problem is further complicated by the possibility that, to some extent, the $J$-band brightening may be caused by a high frequency of unresolved binaries over this spectral type interval \citep{Liu2006}. \citet{Tsuji2003, Tsuji2005} proposed a set of models with dust settling that describes the whole L and T sequence with a parameterization of the cloud deck thickness. They introduced a critical temperature, T$_{\rm cr}$, below which the dust no longer remains in the atmosphere in the form of clouds and is removed from the atmosphere through gravitational settling. They suggest that the $J$-band brightening may be explained by the fact that the L-T sequence is {\it not} necessarily a T$_{\rm eff}$ one but that it must instead be understood as arising from variations of both T$_{\rm eff}$ and T$_{\rm cr}$. Aside from this possibility, it has been suggested that clearings in an otherwise cloudy atmosphere may be responsible for the observed $J$-band brightening \citep{Burgasser2002b}. In such a scenario, where the atmosphere is not uniform, photometric variability could be observed due to the rotation and/or evolution of the cloud pattern.

There have been many observational searches for variability in ultracool dwarfs (M7 and later) in the last few years. Early searches concentrated mostly on L dwarfs in the $I$ band and clearly established that roughly half of these objects display variability on timescales ranging from tens of minutes to weeks, with amplitudes of a few tens of mmag \citep{Tinney1999, Gelino2002, Bailer-Jones2001, Terndrup1999, Martin2001}. Several studies suggest that variability also extends to BDs at the L/T transition. \citet{Enoch2003} reported relatively large $K_{\rm s}$ variability ($\Delta K_{\rm s} \sim 190-420$ mmag with uncertainties of $40-50$ mmag) for 3 transition objects (L7, T1 and T5) with some evidence of periodic ($1-3$ h) modulation. Other groups \citep{Clarke2008, Koen2004, Koen2005} have detected variability for a few objects  at a level of $\sim15$ mmag. Recently, \citet{Goldman2008} described a marginal detection of variability for the T2 dwarf SDSS1254 using near-infrared spectroscopy. Finally, \citet{Morales2006} tentatively detected photometric modulations ($\sim$10 mmag at $4.5 \mu$m) in the thermal infrared for two late-L dwarfs. Overall, while variability seems to be established for L/T BDs, the significance of these detections remains marginal and no clear correlation between photometric bands has been reported.

The T2.5 brown dwarf \hbox{SIMP J013656.57+093347.3} (\citet{Artigau2006}, SIMP0136 hereafter), the second brightest known T dwarf after $\epsilon$ Indi Bab, is an ideal target for studying the intriguing physical properties of BDs straddling the L/T transition. It falls in the middle of the $J$-band brightening and it does not appear to be a binary, down to separations of 0.2$\arcsec$ and $\Delta K=2$ as inferred from Lick AO observations \citep{Goldman2008b}. In this paper, we report on the detection of significant near-infrared photometric variability for this object. The analysis of these observations not only provides compelling evidence for the presence of clouds at the surface of this BD but also offers an attractive explanation for the $J$-band brightening observed in early T dwarfs.

\section{Observations and Data Reduction} \label{sec1}

\begin{deluxetable}{lcccl}
\tablecolumns{5}
\tablecaption{Properties of the reference stars.}
\tablehead{
\colhead{2MASS designation} & \colhead{$J^{\rm a}$} & \colhead{$K_{\rm s}^{\rm a}$} & \colhead{Sep. from} & \colhead{Comment} \\
 & & & \colhead{SIMP0136} &
 }
\startdata
J01365662+0933473& $13.45$ & $12.56$ & &SIMP0136 \\
J01370717+0935071& $13.43$ & $12.54$ & $3.5^{\prime}$ & Comparison star \\
J01364524+0931540 & $11.82$ & $11.54$ & $3.7^{\prime}$ & Reference star \\
J01370290+0937184 & $12.72$ & $12.35$ & $4.3^{\prime}$ & Reference star \\
J01371365+0933327 & $11.85$ & $11.18$ & $4.7^{\prime}$ & Reference star
\enddata
\tablenotetext{a}{2MASS\citep{Skrutskie2006}  point source catalog.}\label{tbl1}
\end{deluxetable}

The $J$-band observations were carried out at the {\it Observatoire du Mont-M\'egantic} (OMM) $1.6$~m telescope with the CPAPIR near-infrared camera \citep{Artigau2004}, featuring a 0.89$\arcsec$ pixel scale and a $30\arcmin\times30\arcmin$ field of view. The observations were obtained on 2008 September 16, 18, 19 and 21, under photometric conditions. As the CPAPIR point-spread function (PSF) would be undersampled under better than 1.8$\arcsec$ seeing, and given that SIMP0136 is a relatively bright target, the camera was purposely defocused to produce $6$-pixel FWHM PSFs. The individual exposure times were $14.8$~s with a single coadd. The data were obtained without dithering and using a guide probe. To further minimize drifts of the field during the sequence due to flexure between CPAPIR and the guide probe in the telescope bonnette, a script was run in real time to send slow ($\sim$ 10 minutes), closed-loop corrections to the guide probe. The results from the first set of observations prompted us to take consecutive $J$ and $K_{\rm s}$-band photometric observations on the night of 2008, December 13. The data were also obtained at the OMM but using the SIMON near-infrared spectro-imager with a $0.46\arcsec$ pixel scale and a $7.8\arcmin$ field of view.  The $J$ and $K_{\rm s}$ sequences lasted $3.0$ and $2.0$ hours respectively with individual exposure times of $29$~s in both bands. The observing strategy was the same as before. For all datasets, a high signal-to-noise (>$10^6$ e$^-/{\rm pixel}$) dome flat was obtained at the end of the night.

All datasets were reduced using the same method. First, the images were dark subtracted and divided by the median-combined dome flat. Then boxes of $50\times50$ pixels centered on SIMP0136, on three nearby reference stars slightly brighter than SIMP0136 and a on comparison star closely matching the $J$ and $K_{\rm s}$ of SIMP0136 (see Table~\ref{tbl1}). The same reference and comparison stars were used for both the CPAPIR and SIMON observations. A handful of bad pixels, none of them falling on the PSFs, were present in the boxes; they were simply interpolated. Aperture photometry was performed using radii of 2.5 times the $50\%$ encircled energy radii of the PSF (typically 5-6 pixels). The sky level was estimated by using the median value of all pixels beyond 1.5 times the photometric aperture radius in each box.

\begin{figure}
\epsscale{1}
\plotone{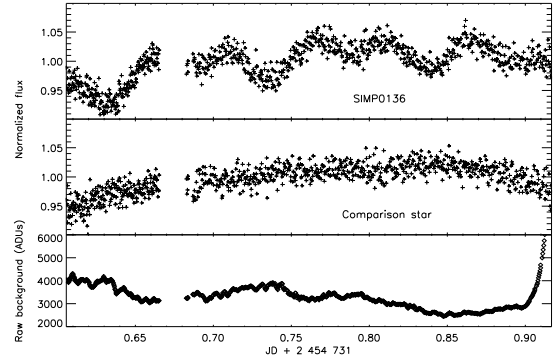}
\caption{Raw light curves for SIMP0136 and the comparison star on September 21, 2008. The modulation is clearly seen in SIMP0136 light curve. The slow modulation in the reference star light curve is seen for all objects and is due to the airmass change during the night ($\sim$1.95 at the beginning and end of the sequence and a minimum of 1.24 at transit).\label{fig5}}
\end{figure}

\begin{figure}
\epsscale{1}
\plotone{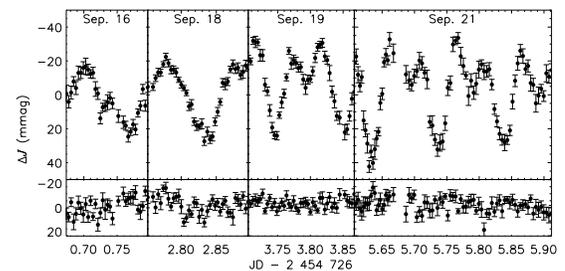}
\caption{The light curves of SIMP0136 for each observing night are shown on the upper panel. The lower panel shows the light curves of the comparison star. The light curves are binned over 5-minute time bins. \label{fig1}}
\end{figure}

The above steps produced raw light curves for the target and each of the three reference stars. The raw light curves of SIMP0136 display a strong modulation unlike anything seen in the raw light curves of the reference or comparison stars (Figure~\ref{fig5}). The raw curves of SIMP0136 and the comparison star were then corrected for systematic effects in the following manner. The light curve of SIMP0136 and the comparison star (shown in Figure~\ref{fig1}) were obtained by dividing them by the mean normalized raw light curves of the 3 reference stars, see Figure~\ref{fig1}. We obtained between 460 and 867 individual images per night and the photometric points shown in Figure~\ref{fig1} are binned over 5-minute intervals (11 to 14 individual measurements). Uncertainties per individual image ranged from 14 to 19~mmag for both SIMP0136 and the comparison star, corresponding to a $\sim5$~mmag uncertainty over a 5-minute interval for both objects. A similar analysis was done for the December 13 data and the results are shown in Figure~\ref{fig2}.

\section{Discussion} \label{sec2}

\begin{figure}
\epsscale{1}
\plotone{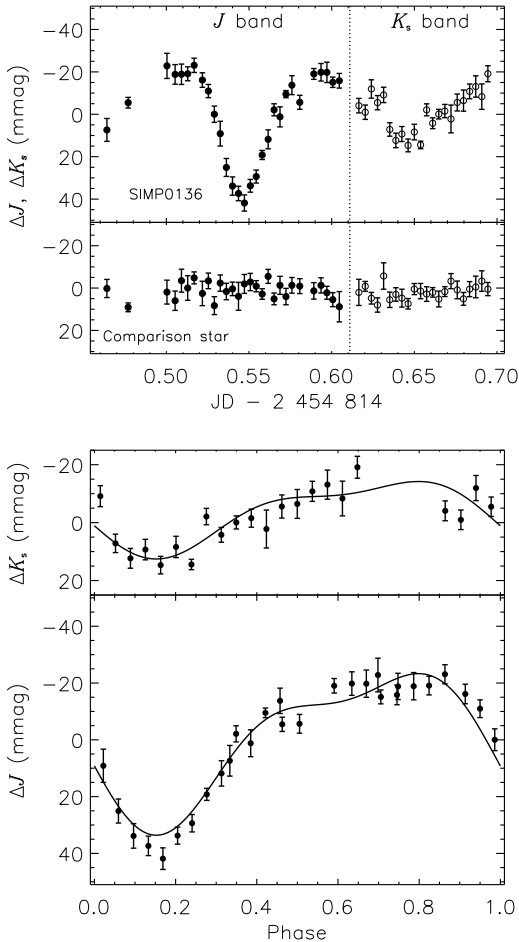}
\caption{$J$ and $K_{\rm s}$ band light curves obtained on December 13, 2008. The upper panel shows the complete light curves of SIMP0136 and the comparison star while the lower one shows the light curves of SIMP0136 folded to a \hbox{2.895~h} period. The modulation observed in September 2008 is still clearly present and observed in both bands. The amplitude of the $K_{\rm s}$ modulation is correlated with the $J$-band one but with an amplitude $0.48\pm0.06$ times smaller. The solid line in the $J$-band phase-folded light curve shows the smoothed curve and the continuous line in the $K_{\rm s}$-band phase-folded light curve is the same line multiplied by $0.48$. The phase-folding relative to the September 2008 observations is arbitrary.\label{fig2}}
\end{figure}

The photometric variability of SIMP0136 is clearly visible from Figure~\ref{fig1}. The variation reached a peak-to-peak amplitude of $\Delta J \sim 50$~mmag. The December data (see Figure~\ref{fig2}) clearly show that the $J$ and $K_{\rm s}$ light curves are correlated with one another, the variations being more pronounced in the former with an amplitude ratio of $\frac{\Delta K_{\rm s}}{\Delta J} = 0.48\pm0.06$. The September 21 light curve, covering 7.5 h, shows a periodic, double-peaked signal. A very similar modulation is also observed on September 19. As a rough estimate of the period can be readily obtained from the September 21 light curve, and as there is an important second harmonic component in the data set, we determined the precise period by fitting a two-harmonic signal using a Levenberg-Marquardt algorithm rather than using a Lomb-Scargle period-search algorithm \citep{Lomb1976, Scargle1982}, which assumes a sinusoidal signal and is better suited for sparsely sampled datasets. We derived a $2.3895\pm0.0005$~h period using either the September 19 and 21 datasets. The evolution of the light curve of SIMP0136 prevents us from deriving a more accurate period as the light curve cannot be folded over several nights. Besides, differential rotation probably renders pointless the attempt of establishing a much more precise period.

A $2.3895$ h period would correspond to a $v \sin i$ of up to $\sim~45$~km/s depending on the object's rotation axis inclination with our line of sight, consistent with the fastest rotators in the \citet{Basri2000} and \citet{Zapatero2006} samples. Only a handful of brown dwarfs have a similar rotation period. The L2 dwarf Kelu-1 has a rotation period of less than 2 hours \citep{Clarke2002}. The young M8.5 dwarf S Ori 45 displays a photometric (in the $J$ and $I$ band) modulation with a $\sim2.6$ hours period but this periodicity was not detected in later observations despite hints of a yet faster (30-45 minutes) modulation \citep{Zapatero2003}.  Strong discrete periodograms peaks have been reported for 2M1145 at periods ranging from 2 to 3.5 hours \citep{Bailer-Jones2001} but the light curve of this object cannot unambiguously be folded to a single period.

\begin{figure}
\epsscale{1}
\plotone{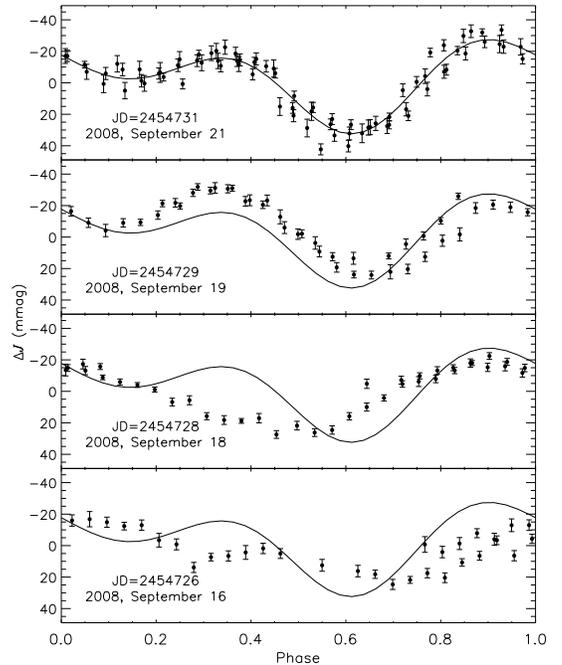}
\caption{Phase-folded light curve of SIMP0136 for each observing night assuming a period of \hbox{2.3895~h}. The continuous line in each epoch's graph represents the smoothed light curve observed on 2008 September 21.  An evolution of the light curve is clearly observed at successive epochs. The double-peaked feature clearly detected on September 19 and 21, and marginally on September 16, has disappeared on September 18.\label{fig3}}
\end{figure}

 The clear evolution of the light curves from night to night in Figure~\ref{fig3} could be due to time-varying clouds or storms in the atmosphere of SIMP0136. In that case, observations over several days can give a time scale for the evolution of those clouds/storms. The evolution of the modulation could also be attributed to a differential rotation. If such an assumption is made, a significant phase difference between the equator and the higher latitudes has to be present within our five-day baseline (here we assume that the December observation is entirely decorrelated). Assuming a (maximum) phase difference of 180$^\circ$ between the equatorial and high-latitude regions of SIMP0136 over the 5-day-long observing sequence would imply a $\sim 1$\% differential rotation. This value is comparable to Jupiter showing a $\sim 0.8$\% differential rotation per period between its equator and high latitudes. It is thus possible that the evolution of the light curves in Figure~\ref{fig3} is the result of differential rotation. This hypothesis will require a much better temporal sampling to be confirmed.

The relative variations at $J$ and $K_{\rm s}$ can be used for constraining the physical processes at play. Using the \citet{Allard2003} and \citet{Burrows2006} models, assuming an effective temperature of $1200$~K, $\log g =5.0$ and solar metallicity, we derived $\frac{\Delta K_{\rm s}}{\Delta J}$ for three different variability scenarios: 1) {\it dusty} clouds on a {\it clear} atmosphere with both phases at the same effective temperature, 2) inhomogeneous surface temperature, for either a {\it dusty} or a {\it clear} atmosphere and 3) {\it dusty} clouds on a {\it clear} atmosphere with both phases at different temperatures. Here {\it dusty} and {\it clear} refer to the {\it settled} and {\it condensed} models of \citet{Allard2003} and to the {\it cloudy} and {\it clear} models of \citet{Burrows2006}. For scenarios 2 and 3 we further considered the models of \citet{Tsuji2003, Tsuji2005}, the case where dust clouds are completely absent from the atmosphere is presented only as a limiting case (called {\it case C} in \citet{Tsuji2003, Tsuji2005}), it assumes a T$_{cr}$ equal to the condensation temperature of the dust grains.

For the first scenario, both sets of models predict that, for an atmosphere of constant temperature, the $J$ and $K_{\rm s}$ band light curves should be anti-correlated (i.e.  $\frac{\Delta K_{\rm s}}{\Delta J} $<$0$). This can easily be understood as the cloud-free models are brighter than the cloud-bearing models at $J$ and inversely at $K_{\rm s}$. Thus if an object evolves from, say, a $50\%$ cloud-free and $50\%$ cloud-bearing atmosphere toward a $51\%$ cloud-free and $49\%$ cloud-bearing atmosphere, it will brighten in $J$ and dim in $K_{\rm s}$.

In the second scenario, the variability is caused by surface temperature inhomogeneities in an entirely {\it clear} or {\it dusty} atmosphere resulting in a time-variation of the disk-averaged temperature. Numerically, this is done by taking the ratio of the relative flux increases in $J$ and $K_{\rm s}$ for a small (i.e. $\Delta T\ll T$) temperature change :

\[ \frac{\Delta K_{\rm s}}{\Delta J} =  \frac{ \left( \frac{\partial F_{K_{\rm s}}}{\partial T} \right) }{\left( \frac{ \partial F_{J}}{\partial T}\right)} \frac{ F_{J} }{ F_{K_{\rm s}} }. \]

 Using all models, we derived a $\frac{\Delta K_{\rm s}}{\Delta J}$ ranging from $1.2$ to $3.4$, the lower and upper limits being set respectively by the {\it cloudy} and {\it settled} models of \citet{Allard2003}. The \citet{Tsuji2003} models using a T$_{\rm cr}=1800$, yield a $\frac{\Delta K_{\rm s}}{\Delta J} \sim 2.0$.


The observed $\frac{\Delta K_{\rm s}}{\Delta J}=0.48\pm0.06$ allows us to exclude the first two variability scenarios. We now consider the third case whereby clouds have a different temperature from the surrounding cloud-free atmosphere. Let $f$ be the cloud filling factor and $\Delta f$ the variation in cloud coverage. We assume that {\it clear} (cloud-free) regions are at a temperature $T = 1200$ K, appropriate for a T2.5 dwarf, and that clouds have a temperature of $T+ \Delta T$. Let $F_{\rm d}[T]$ and $F_{\rm c}[T]$ be the fluxes in the photometric bandpass $m$ as a function of temperature for the {\it dusty} and {\it clear} models. The variability is caused by the variation in cloud coverage ($\Delta f$) as the brown dwarf rotates. It can be shown that the resulting variability $\Delta m$ in a photometric bandpass is 

\[ 
\Delta m  =  2.5 \log \left( \frac{ F_{\rm d}[T+\Delta T] f+F_{\rm c}[T] (1-f)  }{   F_{\rm d}[T+\Delta T] (f+\Delta f)+F_{\rm c}[T] (1-f-\Delta f) } \right). \]

\begin{figure}
\epsscale{1}
\plotone{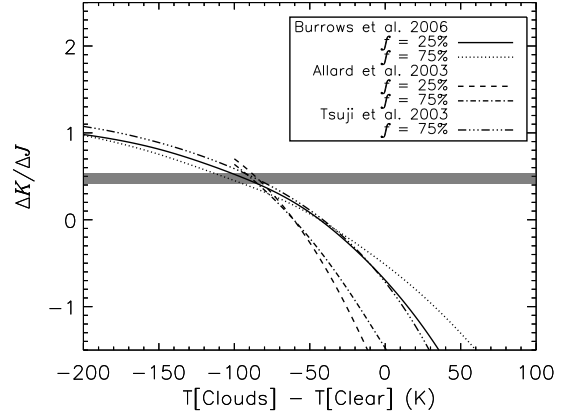}
\caption{$K_{\rm s}$ versus $J$ band variability as a function of the temperature difference between the cloudy and cloud-free atmosphere patches. The horizontal line corresponds to the observed $\frac{\Delta K_{\rm s}}{\Delta J}$ and the thickness of the line to the uncertainties on that measurement. The \citet{Allard2003} models do not extend below $\Delta T = -100$ K as the {\it settled} models have only been computed down to 1100 K. All models assume a temperature of 1200 K for the cloud-free atmosphere, $\log g = 5.0$ and solar metallicity.\label{fig4}}
\end{figure}

The $\frac{\Delta K_{\rm s}}{\Delta J}$ ratios for all models and different values of $f$ and $\Delta T$ are shown in Figure~\ref{fig4}. As shown in this figure, the resulting $\frac{\Delta K_{\rm s}}{\Delta J}$ varies only slightly as a function of the assumed filling factor for clouds versus cloud-free patches. Also, the overall behavior between the \citet{Allard2003} and \citet{Burrows2006} models is consistent. The observed $\frac{\Delta K_{\rm s}}{\Delta J} = 0.48\pm0.06$ can be reproduced by clouds between 80 to 110 K colder than the cloud-free atmosphere. The results are mildly dependent on the temperature of the {\it clear} regions; a $T$=1300 K would increase the derived $\Delta T$ to values ranging from 125 to 150 K. This variability scenario also gives a $\frac{\Delta f}{\Delta J}$ ratio allowing for a direct conversion of the light curve amplitude ($\Delta J$) into an estimate of the variation of the cloud filling factor ($\Delta f$). The derived $\frac{\Delta f}{\Delta J}$ range from 1.5 to 2.3 depending on the assumed filling factor $f$ (between 0.25 and 0.75). This implies that for a $50$ mmag variability amplitude in $J$, the patches involved in SIMP0136 variability cover between $7\%$ and $16\%$ of the visible disk. This does not constrain the overall filling factor of the {\it dusty} versus {\it clear} regions, only time-differences.

In the \citet{Tsuji2003, Tsuji2005} models, the {\it case C} (equivalent to the {\it clear} models) is not expected to cover a large fraction of the surface, we therefore examined only the case $f=0.75$ case where $75\%$ of the surface would be covered by the T$_{\rm cr}=1800$ model (equivalent to the {\it dusty} models). The observed $\frac{\Delta K_{\rm s}}{\Delta J}$ implies a $\sim100$K temperature difference between the {\it case C} and cloud-bearing regions, consistent with the predictions of the \citet{Allard2003} and \citet{Burrows2006} models.

Beside atmospheric phenomena, periodic modulation at a few percent level could easily be introduced by the presence of a transiting companion with a radius of 2-3 R$_{\earth}$ or by grazing transits in a roughly equal-sized binary. However, the observed night-to-night evolution of the light curves strongly disfavors any explanation of the variability in terms of transits/grazing transits. The expected light curves would be very stable from period to period and evolve extremely slowly through the evolution of the orbital elements. Also, any transit scenario would imply an out-of-transit flat-top light-curve that is not observed. Furthermore, transit light curves show little wavelength dependence (apart from limb darkening effects) and one would expect a $\frac{\Delta K_{\rm s}}{\Delta J} \sim 1$ regardless of the geometry of the system.

\section{Concluding remarks} \label{sec3}
The observed variability of SIMP0136 suggests that atmosphere models attempting to describe its physical properties using a single set of parameters (dust settling rate, temperature, gravity, etc) are bound to be incomplete as at least two different types of regions must be present on its visible disk to produce the observed variability. The idea that early-T dwarfs could be described as objects with more than one set of physical properties and arbitrary filling factors is not new \citep{Burgasser2002b}, but the detected variability and its interpretation in terms of colder clouds in a warmer atmosphere provides empirical physical constraints of these cloudy and cloud-free regions.

It has been proposed previously that the $J$-band brightening may result mainly from a decrease in the fractional coverage of dust clouds against warmer, {\it clear} regions of the brown dwarfs' atmospheres as they progress from spectral type L to T. Indeed, a brightening in $J$ is a natural outcome of this as warmer {\it clear} regions are brighter than cooler {\it dusty} regions in $J$. Our observations not only bring empirical evidence for this scenario, but also provide an estimate of the temperature difference between the {\it cloudy} and {\it clear} regions of the atmosphere. The difference of $\sim$100 K suggested by our results would lead to a brightening of $0.64$ mag in $J$ from a completely {\it cloudy} to a completely {\it clear} atmosphere, assuming no temperature change for each respective model; this amplitude is similar to that of the observed $J$-band brightening. In reality, the L/T transition is likely accompanied by a mild change in the temperature of both types of atmosphere -  {\it cloudy} and {\it clear} - but the overall behavior would be similar. The exact shape of the $J$-band brightening could potentially be reproduced by adjusting, as a function of spectral type, both the temperature of the {\it clear} regions and the difference of temperature between the {\it clear} and {\it cloudy} patches. 

Considering that most searches for variability in early-T dwarfs resulted in non-detections or marginal detections despite sensitivities sufficient to readily detect variability at the level observed for SIMP0136, the following question naturally arises : is SIMP0136 an oddball object with unusual physical properties ? or is it representative of early-Ts ? The non-detection of variability in other early-Ts does not disprove the {\it cloudy} versus {\it clear} hypothesis: cloud structure on these objects could be either too small to produce significant photometric variability or distributed in bands without longitudinal inhomogeneities. The near-infrared photometric properties of SIMP0136 are typical of early-T dwarfs (See Figure 2 in \citet{Artigau2006}) and its proper motion is consistent with that of a thin disk object; it is therefore not expected to have unusually low metallicity and/or high surface gravity. The only slightly unusual property of SIMP0136, besides its photometric variability, is its relatively rapid rotation. Determining whether rapidly rotating brown dwarfs are more likely to feature larger-scale atmospheric structures would require global circulation models; this could be verified observationnaly by a variability survey specifically targeting rapid rotators.

The physical interpretation of the mechanisms at play is strongly dependent on the accuracy of current atmospheric models and their handling of dust settling; it is nevertheless encouraging that three major sets of models give similar temperature differences and surface coverage. The variability model proposed here leads to a number of readily testable predictions that will provide a unique input for modelers. Photometric variability in other bands is expected; $\frac{\Delta H}{\Delta J}$ should closely match $\frac{\Delta K_{\rm s}}{\Delta J}$, while $\frac{\Delta Y}{\Delta J}$ should be $\sim 0.8$. Obtaining an accurate measurement of the $\frac{\Delta i}{\Delta J}$ would be of particular interest as this ratio is strongly dependent on the temperature difference between the {\it cloudy} and {\it clear} regions, ranging from $+1.5$ to $-2$ over the $\Delta T = -200 $ to $+40$~K interval. Time-resolved spectroscopy should reveal a number of variable spectroscopic features, especially in the far-red and $J$ bands. In particular, variability is expected for the depth and shape of the 0.78 $\mu$m and $J$-band K{\sc I} doublets, the continuum slope between 1.17 and 1.22~$\mu$m, and the strength of the $\sim 0.86$~$\mu$m CrH and FeH bands.

\acknowledgments

We would like to thank our referee, Victor B\'ejar, for useful suggestions and comments that significantly improved our manuscript. We thank Philippe Delorme for his thoughtful comments on the manuscripts and C\'eline Reyl\'e for her help on determining the likely membership of SIMP0136 to the thin disk. EA is supported by the Gemini Observatory, which is operated by the Association of Universities for Research in Astronomy, Inc., on behalf of the international Gemini partnership of Argentina, Australia, Brazil, Canada, Chile, the United Kingdom, and the United States of America. DL is supported via a postdoctoral fellowship from the Fonds Qu\'eb\'ecois de la Recherche sur la Nature et les Technologies. RD is financially supported via a grant from the Natural Sciences and Engineering Research Council of Canada.

\bibliographystyle{apj}


\clearpage


\end{document}